\documentclass[prl,twocolumn,preprintnumbers,showpacs]{revtex4-1}
\usepackage{amsmath, mathrsfs, amssymb,amsfonts,amsthm,graphicx, epsf, dcolumn, yfonts}
\usepackage[hyperfootnotes=true]{hyperref}
\usepackage{subfigure}
\usepackage{color}
\usepackage{slashed}
\usepackage{setspace}
\usepackage{cancel}
\usepackage{wasysym}
\pdfoutput=1
\newcommand{\be}{\begin{equation}}
\newcommand{\ee}{\end{equation}}
\newcommand{\bea}{\begin{eqnarray}}
\newcommand{\eea}{\end{eqnarray}}

\def\pp{\partial}

\newcommand{\nn}{\nonumber \\}

\def\lab{\label}

\def\cA{{\cal A}}  
  
 \def\cH{{\cal H}} 
  \def\cL{{\cal L}}
\def\cM{{\cal M}}

\newcommand{\prt}[1]{{\left( {#1} \right)}}
\newcommand{\prtt}[1]{{\left[ {#1} \right]}}

\def\IR{\relax{\rm I\kern-.18em R}}

\def\pp{\partial}

\def \tx  { {\tilde {x }}}
\def \tphi  { {\tilde {\phi }}}

\def\IR{\relax{\rm I\kern-.18em R}}
\def\IL{\relax{\rm I\kern-.18em L}}

\def\inv{^{\raise.15ex\hbox{${\scriptscriptstyle -}$}\kern-.05em 1}}

\def\cM{{\cal M}}

\def\cL{{\cal L}}

\def\cA{{\cal A}}

\def\bea{\begin{eqnarray}}
\def\eea{\end{eqnarray}}
\newcommand{\eq}[1]{(\ref{#1})}
\def\nn{\nonumber}

\newcommand{\la}[1]{\label{#1}}

\def\a{\alpha}      
\def\b{\beta}       
\def\g{\gamma}    
\def\d{\delta}    
\def\e{\epsilon}

\def\k{\kappa}
\def\l{\lambda} 
\def\m{\mu}

\def\r{\rho}
\def\s{\sigma}  
\def\t{\tau}

\def \dz {\dot{z}}

\def \dx {\dot{x}}

\definecolor{markcolor2}{rgb}{1,0,0}

\definecolor{markcolor3}{rgb}{0,1,0}

\newcommand{\diff}{\mathrm{d}}

\renewcommand*{\thefootnote}{\fnsymbol{footnote}}

\begin{document}

\title{Analytic Non-Integrability and S-Matrix Factorization}
\author{Dimitrios Giataganas}
\email{dgiataganas@phys.uoa.gr}

\affiliation{\vspace{1mm}
Department of Physics, University of Athens, 15771 Athens, Greece.\\
and\\
Albert Einstein Center for Fundamental Physics, Institute for Theoretical Physics, University of Bern, Sidlerstrasse 5, 3012 Bern, Switzerland.}

\begin{abstract}\vspace{-2mm}

We formulate an equivalence between the 2-dim $\sigma$-model spectrum expanded on a non-trivial massive vacuum and a classical particle Hamiltonian with variable mass and potential. By considering methods of analytic Galoisian non-integrability on appropriate geodesics of the Hamiltonian system we algebraically constrain the particle masses at fixed time, such that integrability is allowed. Through our equivalence this explicitly constrains the masses of the excited spectrum of the dual 2-dim theory in such a way to imply the S-matrix factorization and no particle production.  In particular, the integrability of the classical particle system, implies the factorization of the S-matrix in the dual quantum 2-dim theory. Our proposal provides also non-trivial evidence without any assumptions, on the connection between integrability and S-matrix factorization for large class of theories with interactions that break Lorentz invariance.

\end{abstract}

\renewcommand*{\thefootnote}{\arabic{footnote}}
\setcounter{footnote}{0}


\maketitle


\noindent \textbf{1. Introduction.}
\lab{sec::intro}
Integrable field theories have been known to have pivotal role on understanding physical structures such as the bosonization and the properties of the factorization of scattering \cite{Beisert:2010jr,Bombardelli:2016rwb}. On the other hand integrable field theories are very few and hard to find in nature, their population is usually resembled by a small region of islands in a large ocean of water. The standard study on the integrable structures is done by the construction of Lax pair for the relevant sigma model, a complicated task with no standard methodology that can be based on the symmetries or properties of the theory. It has been practically more applicable in theories that admit a deformation of an integrable parent theories where one keeps track of the Lax pair deformation, with several modern and older applications including \cite{Frolov:2005dj,Beisert:2010jr,Klimcik:2019kkf,Orlando:2019rjg}.

Due to these reasons there is an extensive research of alternative methods for the study of integrability. A natural development to this direction are formalisms focusing on the necessary conditions for the existence of integrability. Such are the  methods of analytic non-integrability which are primarily based on the connection of differential Galois theory with the Hamiltonian equations and the way that the integrable systems behave under certain fluctuations \cite{Fomenko,MoralesRuiz,Goriely0,morales1994ps}. The idea can be implemented by the Kovacic algorithm where the question of (non-)integrability boils down to simpler algebraic statements \cite{Kovacic1}.
The recent years the method has been found several application in holographic models, initiated with the string dynamics in black hole environments \cite{Zayas:2010fs}, while the literature so far focuses on the application of the method in the different theories with aim to classify their integrability status including \cite{Basu:2011fw,Stepanchuk:2012xi,Giataganas:2013dha,Chervonyi:2013eja,Giataganas:2014hma,Asano:2015eha, Ishii:2016rlk, Hashimoto:2016wme, Nunez:2018qcj,Nunez:2018ags,Giataganas:2017guj,Akutagawa:2019awh}.

An independent criterium of integrability is generated by the classical S-matrix on a 2-dim massive theory. The higher conserved charges imply equal sets of masses and momenta before and after the collision and the absence of particle production. The locality and causality impose factorization of the $n\rightarrow n$ amplitudes into products of $2\rightarrow 2$ ones. Therefore, integrability simply requires that the $S$-matrix factorizes and satisfies the conditions of the no particle production in Lorentz invariant theories with massive spectrum \cite{Zamolodchikovf,Shankar:1977cm,Dorey:1996gd}.  However, even for theories where Lorentz invariance is broken by the interactions and a massless spectrum exist, which still may  not contribute to the amplitudes, the relation between integrability and factorization scattering used to be viewed as a logical assumption \cite{Hoare:2018jim,Wulff:2019tzh}. Nevertheless, formal proofs exist only for theories that preserve the Lorentz invariance, for example \cite{Parke:1980ki}. Our current work provides non-trivial evidence for a large class of theories for the validity of this statement. While the S-matrix factorization  has been related to integrability since the very early studies, only recently it has been used as an application to classify non-integrable structures \cite{Wulff:2019tzh,Wulff:2017lxh,Wulff:2017vhv,Wulff:2017hzy}.

In this paper we initiate a study of combining the ideas of the two methods, by applying analytic non-integrability methods on configurations that contain the non-trivial vacuum of the worldsheet S-matrix. In particular, in the formalism we develop we show that the $\sigma$-model massive field excitations around a non-trivial vacuum are related to dynamics of classical particle systems of variable mass with a potential that depends on the details of the equivalent 2-dim theory and the vacuum chosen. On the resulting effective particle Hamiltonian that the string system is reduced, the analytic  Galoisian methods provide a non-trivial algebraic relation on the particle's mass parameter, imposing strong fine-tuning to allow the integrability of the classical system.

An especially interesting question concerns the interpretation of the integrability particle constrains, when mapped on the 2-dim world-sheet theory. By computing the spectrum excitations on the chosen vacuum and the relevant amplitudes we find that the classical particle integrability conditions, imply the no-particle production and the factorization of the S-matrix of the quantum 2-dim $\sigma$-model. Our proposal therefore opens a new ground for applications on the non-integrability techniques on the S-matrix vacua, which goes beyond the classification techniques.

Moreover, the relation between the factorization of a S-matrix with broken Lorentz symmetry by interactions and the integrability of the $\s$-model is a priori unclear and in many cases used to be an assumption. We find systematic evidence of this connection for a large class of theories and validate therefore the assumptions of the existing literature. Our methodology may provide a ground to develop a formal proof of this conjecture by working on the dual particle picture side.

\noindent \textbf{2. The setup.}
\lab{sec:setup}
A generic  background of $d$ space-time dimensions with Minkowski signature and with $n$ cyclic coordinates is described by
\bea\nn
ds^2=g_{ii} \diff x^i \diff x^i+2 g_{ij} \diff x^i \diff x^j~,
\eea
where $i<j$ and $i,j=1,\ldots,d$ and the metric fields are functions of the non-cyclic coordinates. We assign the indices $i\ge d-n$ to label the cyclic angles, on which the metric elements are independent. The Polyakov Lagrangian $\cL$ is given by the following integrand
\be\nn
S=\int d\s d\t g_{ii} \prt{ x^{i\prime 2}-\dot{x}^{i 2}}+2 g_{ij} \prt{x^{i\prime} x^{j\prime}- \dot{x}^i\dot{x}^j}~,
\ee
where the dotted and primed derivatives are with respect to the worldsheet coordinates  $\prt{\t,\s}$.

The equations of motion for the non-cyclic angles $\a^i:=x^i$ with $i<d-n$ are
\be\label{eomnocycle}
\partial_{\a^i} \cL +2 \partial_\t \prt{g_{ii}\dot{\a}^i+g_{ij}\dot{\a}^j}-2 \partial_\s \prt{g_{ii}\a^{i}{}'+g_{ij}\a^{j}{}'}=0~,
\ee
where still $i<j$ and here the $i$ index is not summed since it labels the field of the corresponding equation of motion. The cyclic angles $\phi^i:=x^i$ for $i\ge d-n$ have simpler equations of motion
\be\label{eomcycle}
\partial_\t \prt{g_{i i}\dot{\phi_i}+g_{ij}\dot{\phi_j}}- \partial_\s \prt{g_{i i}\phi_i'+g_{ij}\phi_j'}=0~,
\ee
where the index $i$ is not summed. The Virasoro constraints take the compact form
$g_{ii} \dot{x}^i x'^{i}+g_{ij} \dot{x}^i x'^{j}= 0,$ and 
$\cL_{+}=0,$
where all the indices sum and $\cL_{+}$ is defined from the Lagrangian density expression by flipping the minus signs in front of the kinetic terms.

\begin{figure}[t!]
\includegraphics[angle=0,width=0.5\textwidth]{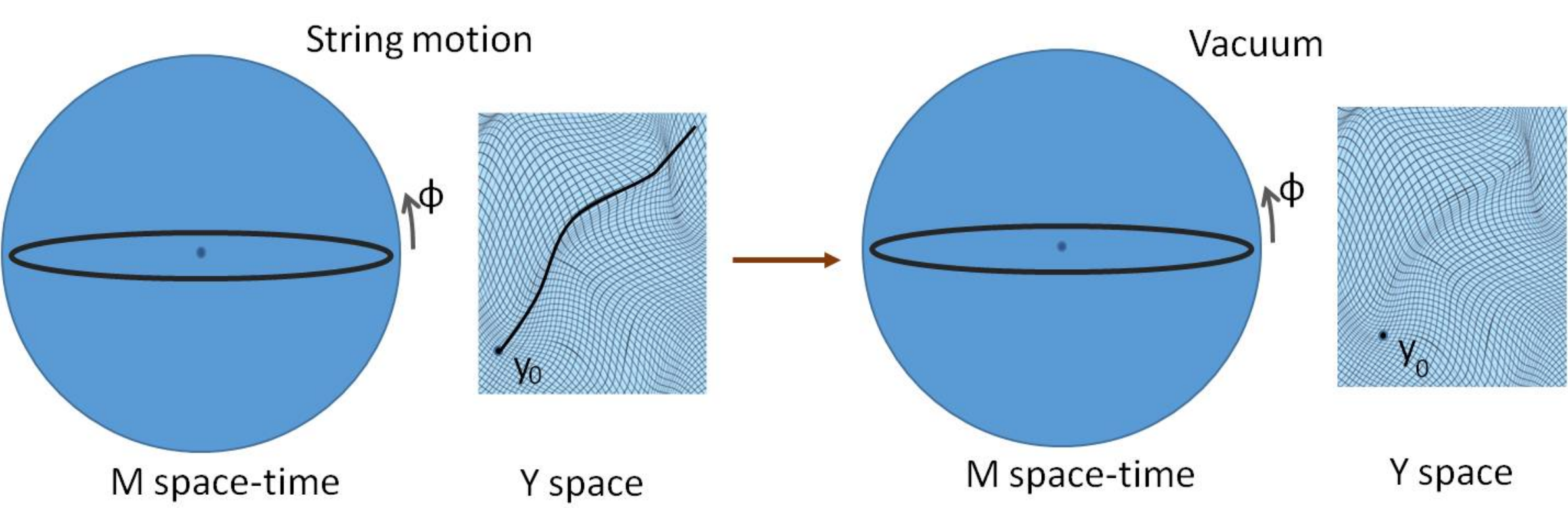}
\caption{\small An example of a string motion in the curved spacetime, to be mapped to the particle classical Hamiltonian. The non-trivial vacuum for the worldsheet scattering is the localized string configuration at $y_0$.
\label{fig:stringvac}}
\end{figure}

\noindent \textbf{3. The Implementation of the Formulation.}
\lab{sec::setup}
In this section we schematically describe the method and the mapping of the particle-string system. We consider a holographic Lorentz invariant geometry, with a metric written in the diagonal form
\be\la{metric1}
ds^2=  ds^2_{M}(\textbf{x},y)   + \diff s^2_Y(\textbf{y})~. 
\ee
The space-time $\cM$ has a boundary and $Y$ is an internal space. The geometry is parametrized by $\mathbf{x}$ and $\mathbf{y}$ respectively.   Let us denote the non-trivial classic vacuum which we expand on for the study of the world-sheet S-matrix to be parametrized by $\{\mathbf{\tilde{x}}(\t,\s),y_0\}$  
and could be for example a Gubser-Klebanov-Polyakov-type (GKP) of string, or another string solution. One of the key ingredients for the equivalence we propose, requires to find the appropriate extended configuration $\{ \mathbf{ x }(\t,\s), y(\t,\s)\}$ which localizes consistently to the vacuum solution at $y_0$, as in Fig. \ref{fig:stringvac}, and solves the Virasoro and equations of motion \eq{eomnocycle} and \eq{eomcycle}.

\begin{figure}[t!]
\includegraphics[angle=0,width=0.5\textwidth]{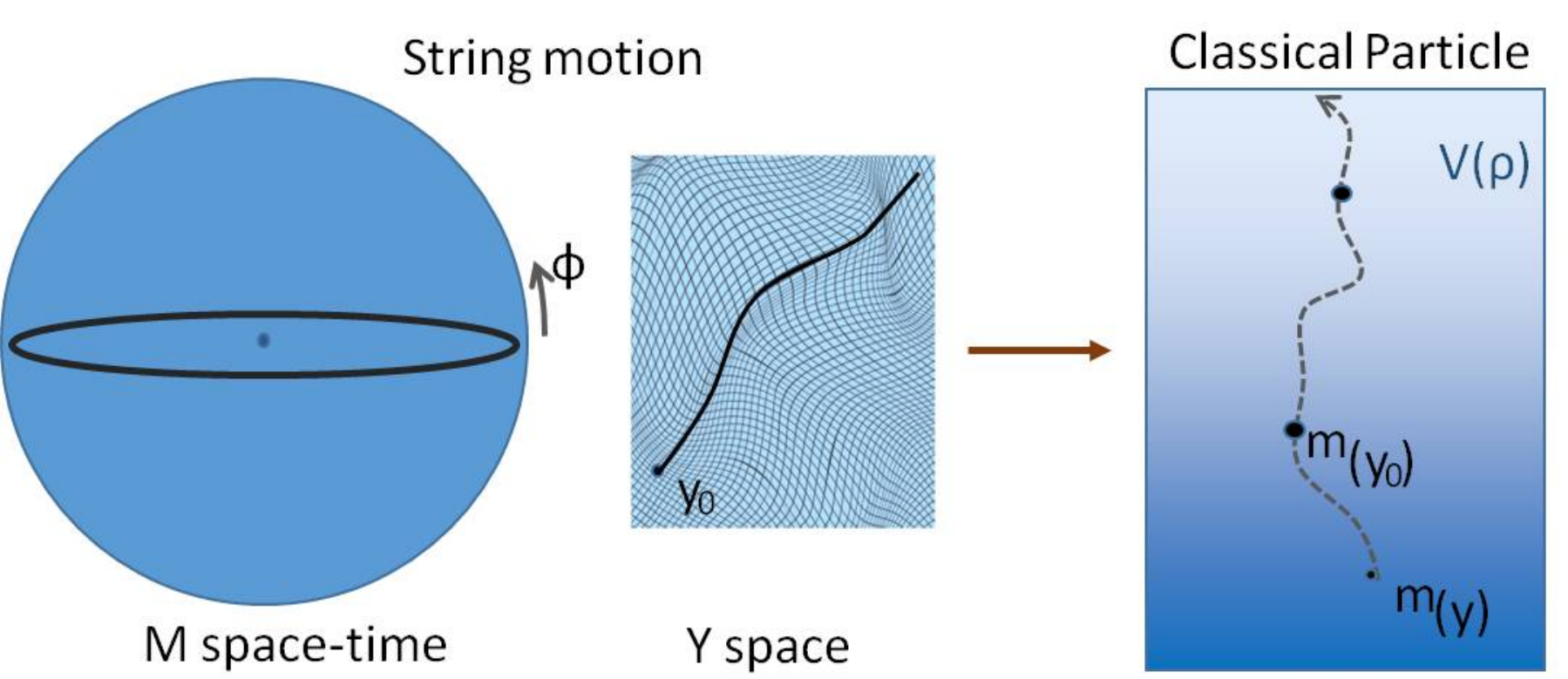}
\caption{\small A string motion in the curved spacetime and its equivalence with a particle of variable mass
in an effective potential $V(\rho)$. The non-trivial vacuum for the worldsheet scattering is the localized configuration at $y_0$. The properties of the mass of the particle at this point, are related to the mass of the excited fields of the 2-dim $\sigma$-model.
\label{fig:stringp}}
\end{figure}

Once we have chosen the vacuum and its appropriate string configuration that localizes to it, we reduce the full system of equations consistently to an effective Hamiltonian of a particle with non-trivial potential, as in  Fig. \ref{fig:stringp}, using the equations of section 2. The Hamiltonian can be brought to the form
\be
\cH= \frac{p_x^2}{2 m(y)}+p_y^2 +V\prt{\mathbf{x},y}~,
\ee
where through the kinetic terms we can read the effective mass $m(y)$ of the particle, related to the corresponding background metric fields \eq{metric1} \footnote{In general we can have a non-trivial mass in $y$-kinetic term, but this does not affect the conclusions of this section. In most cases it can be rescaled to the presented form using the symmetries of the theory. Notice that we will mainly demonstrate later the GKP-types of string, but the particle-string equivalence is valid but more involved for other type of strings.}. The origin of the kinetic terms in general is due to motion on the non-cyclic coordinates and of the potential due to the cyclic ones.  At the point $y_0$, where we recover the vacuum, our prescription is to perform transverse fluctuations on the solution as $\d y=y_0+\eta(\s)$. The eq. \eq{eomnocycle} of $y$ using the rest equations, gives a second order homogeneous differential equation for $\eta$ of the form
\be\la{nvesample}
\eta''(\s)+ h_1(\tilde{x}(\s)) \eta'(\s)+ \pp^2_y m(y_0) ~ h_2(\tilde{x}(\s)) \eta(\s)=0~.
\ee
This is the Normal Variational Equation (NVE) in the appropriate form necessary for the application of the Kovacic algorithm.   The functions $h_1$ and $h_2$ depend on the effective potential $V$ in the particle description, or equivalently on the geometry of the sigma model in the dual string picture. The parameter $m(y_0)$ is a constant at the point where the vacuum in string picture is recovered.

The differential Galois group relates the absence of integrability with the absence of Liouvillian solutions of the NVE. Therefore, we can obtain the non-integrability constrains on the second derivative of the effective mass of the particle $m(y_0)$ to restrict it to a set of values $\cA$, such that only for these values the integrability of the classic system may exist.

Due to the properties of the initial 2-dim string configuration,  the effective particle mass is related to the mass of a field in the spectrum of the world-sheet excitations on the vacuum! To confirm this statement we study the world-sheet scattering where we consider the bosonic string sigma model on the 2-dim vacuum $\{\mathbf{\tilde{x}}(t,s),y_0\}$. Instead of fluctuating the extended string solutions around the vacuum, as in the method described above, here we expand the sigma model. The spectrum is read from the quadratic part of the fluctuation Lagrangian which takes the form
\be
L_2=\prt{\pp x_i}^2+ m_i x_i^2+ \prt{\pp y}^2+ \pp^2_y m(y_0) y^2~,
\ee
where $m_i$ are constants depending on vacuum and the geometry, while the presence of the second derivative of $m(y)$ can be understood by the same order expansion of the Lagrangian. It is the same derivative that appears in the integrability analysis of the particle picture \eq{nvesample}. The requirement of the no particle production and the factorization of the amplitudes requires the ones with unequal incoming and outgoing masses to vanish, constraining therefore the masses of the spectrum $\prt{\sqrt{m_i},\sqrt{\pp^2_y m(y_0)}}$. The quadratic, cubic and quartic order Lagrangians provide the propagators and the vertices to analyze and compute the $2\rightarrow 2$ scattering at tree level and the higher loop corrections to the two-point function.

According to the equivalence we have described between the particle and string picture, the requirement of (non-)integrability in the particle classic system, constrains the spectrum in the dual string side such that the S-matrix factorization in the quantum 2-dim theory occurs!

\noindent \textbf{4. The string-particle mapping on the GKP vacuum case.}
\lab{sec:se3}
Let us demonstrate explicitly the above generic formalism, by considering as an example a background with generic warp factors as
\be\la{metric}
ds^2=g_\a(y) \diff s^2_{AdS}+ g_\b(y) \diff s^2_Y~,
\ee
where the $\mathbf{y}$ coordinates parametrize the internal space $Y$. By a coordinate transformation we could absorb the second warp factor, but let us work initially on the most generic case so our results are directly applicable to any background and we will simplify the presentation later on.
We parametrize the AdS as
\be
ds^2= -\cosh^2 \r~ \diff t^2+ d\r^2+\sinh^2 \r ~\diff\phi^2~,
\ee
where $\r$ is the radial coordinate, $\phi$ is cyclic coordinate and $t$ the time. Higher dimension $AdS$ spaces can be considered without loss of generality, where the $\diff\phi$ is replaced by the $d-2$ metric of the round sphere.

The appropriate string configuration is parametrized on the conformal space by $t=c \t~,~ \phi=c \omega t~,$ with a rigid rotation on the holographic direction $\r(\s)$ and the internal space direction $y(\s)$.

The equations of motion for our fields are obtained by a direct application of \eq{eomnocycle} and \eq{eomcycle}
and are consistent with the Virasoro constraints.
The system is reduced to the particle Hamiltonian system where the role of the time is played by the parameter $\s$ parametrizing the string length.  The effective Hamiltonian in terms of the conjugate momenta $p_y$ and $p_\r$ is given by
\be\la{hamilt}
H_{eff}=\frac{p_\r^2}{4 g_\a }+ \frac{p_y^2}{4 g_\b } + c^2 g_\a \prt{-\cosh^2 \r(\s)+\omega^2 \sinh^2  \r(\s)},
\ee
and is constrained to zero by the Virasoro constraint. The effective potential $V$ carries information of the theory and the string solution, and originates from the cyclic coordinates.

We localize our rotating string around an appropriate point $y_0$, which we set it to 0. As long as the background condition
\be\la{cond_y}
\pp_y g_\a(0)=0~
\ee
holds, the equations of motion are solved consistently for a folded string rigidly rotating, the GKP string \cite{Gubser:2002tv}. The solution for $\r(\s)$ takes the form of the Jacobi amplitude
$ \r(\s)=i am( i c \s,1-\omega^2)~,$
which is the inverse of the incomplete elliptic integral of first kind. The constant $c$ may be adjusted to fix the period of $\s$ to a desirable value, for example $2\pi$. The length of the string depend on the value of the parameter $\omega$. For $\omega$ close to the unit the string is large, while for $\omega=1$ it is infinite. For large $\omega$ the string is short and  the solution can be approximated to the spinning string in flat space, producing the string Regge trajectory. In fact the regular string is a one-soliton sinh-Gordon solution while the long string limit corresponds to a two-soliton configuration.

The NVE is obtained at the point where the string solution is localized in the internal space and recovers the vacuum.  The variation is therefore introduced in the classical particle system as $\d y(\s)=0+\eta(\s)$ and after the appropriate manipulation of the system of equations, the linearized NVE turns out to be equal to
\be\la{nvegkp01}
 \eta''(\s)- g_\a''(0) c^2 \prt{ \cosh^2 \r(\s)-\omega^2 \sinh^2  \r(\s)}\eta(\s)=0~,
\ee
where the derivative on $g_\a$ is with respect to $y$, a convention we adopt for the rest of the paper, and we have used the requirement \eq{cond_y}, necessary  for the existence of the GKP solution in our theory \footnote{We have rescaled \eq{metric} such that $g_\b(0)=1$ to simplify the presentation. Otherwise the ratio $g_\a''(0)/g_\b(0)$ would appear instead the numerator itself.}. We immediately notice that the requirement of integrability can constrain only the second derivative of the  warp factor $g_\a(y)$ at $y=0$. Does the application of the differential Galois group analysis on the NVE \eq{nvegkp01} provides an algebraic analytic condition, fine-tuning the acceleration of change of mass $g_\a''(0)$ of the classical system to values that allow integrability? The answer is positive and this is one additional key ingredient of this work.

\noindent \textbf{5. The Galoisian Non-Integrability on the Particle System.}
\lab{sec:s4}
The rational NVE equation is obtained with the change of variable $z:= \cosh^2 \r(\s)-\omega^2 \sinh^2  \r(\s)~,$ where we identify the resulting equation as the linear general Heun differential equation
\bea \la{heunge}
&&\eta''(z)+\k(z)\eta'(z)+\l(z)\eta(z)=0~,\\\nn
&&\k(z):=\frac{\g}{z}+\frac{\delta}{z-1}+\frac{\e}{z-c_0}~,\quad \l(z):=\frac{\tilde{\a}\b z-q}{z\prt{z-1}\prt{z-c_0}}~,
\eea
which has four regular singular points $\prt{0,1,c_0,\infty}$ on the Riemann sphere  when the condition
$\tilde{\a}+\b-\g-\delta  -\e+1=0 ~$
holds. The transformed NVE \eq{nvegkp01} is identified by $\tilde{\a}=1,~ \g=\delta=\e=\frac{1}{2},~ c_0=\omega^2,~ q=0,~ \b=-g_\a''(0)/4$.

By transforming the Heun equation to a Riccati and using the differential Galois group analysis as described in the appendix A, we apply analytically the Kovacic algorithm to obtain the values of the parameters for which the differential equation has Liouvillian solutions
\be\la{criteria}
\pm\prt{\g+\d+\e+\tilde{\a}-\b}-1 = 2 n  ~,
\qquad n\in\mathbb{Z}~.
\ee
What follows is a generic statement: the integrability of a particle system which produce an NVE of the form \eq{heunge} can not be excluded only for the combination of the parameters that satisfy the above condition.

Therefore the Liouvillian criterium \eq{criteria} applied to our NVE on the Riemann sphere gives the constrain
\be\la{set}
g_\a''(0)=n~, \quad n= \{0,~ 2\}~.
\ee
where the null value corresponds to a trivial linear solution as it can be seen from \eq{nvegkp01}. With the given identification, the effective particle Hamiltonian can be integrable for the values of $g_\a''(0)$ of the equation \eq{set}, which correspond to the acceleration of the change of the particle's mass  \footnote{ We could generalize further to $\tilde{\a}=k~,$ and $\b=-g_a''(0)/(4 k)~,$ with $k\in \mathbb{R}$, to get the condition $g_a''(0)=2 n(2 n -1)~, \quad n\in \mathbb{Z}~$. Then the negative values and the rest of positive are discarded by natural requirements, since they are related through our equivalence to the massive string excitations to end up again with \eq{set}.}.

\noindent \textbf{6. Non-Integrability with Factorized Scattering for the GKP vacuum.}
\lab{sec:s5}
The vacuum of the theory we demonstrate our formalism is the reduced $y_0=0$ configuration of the previous section, the GKP string.
We consider the $AdS$ space of  arbitrary dimension in Poincare coordinates $(x^\m,z)$ with boundary at $z=0$. The bosonic part of the Euclidean action in the light-cone gauge is \footnote{We rescale $g_\b(y)=1$ in this section for convenience in the presentation.}
\be
\cL_E=g_\a(y) \prt{\dz^2+|\dx|^2+\frac{1}{z^4}\prt{z'^2+ |x'|^2}}+z^2 \dot{y}^2+\frac{y'^2}{z^2}~,
\ee
where the condition $\sqrt{-\g} \g^{\a\b}=\mbox{diag}\prt{-z^2,z^{-2}}$ has been imposed on the world-sheet metric. The equations of motion for generic world-sheet dependence on $\prt{\t,\s}$ are obtained following the general derivation of section 2. When the derivative of the warp factor of the metric satisfies the condition \eq{cond_y}, i.e. $g_\a'(0)=0$, we see that the system admits the generalized null cusp \cite{Kruczenski:2002fb} solution
which ends on the null cusp at the boundary of the space,
as in \cite{Giombi:2009gd,Giombi:2010fa}. Following \cite{Wulff:2019tzh,Bianchi:2015iza}, the fluctuations around the null cusp with
$z = \sqrt{\t/\s} \tilde{z}~,$  $\tilde{z}\equiv e^{\tilde{\phi}}~,$
$x=\sqrt{\t/\s} \tilde{x}~,$
and world-sheet coordinate change $\tilde{\t}=2 \ln \t$ and $\tilde{\s}=2\ln \s$ that makes the induced world-sheet conformally flat metric, gives the quadratic action
\be
\cL_2= (\pp_\a \tilde{\phi})^2+\prt{2\tilde{\phi}}^2+ (\pp_\a \tilde{x})^2+\prt{\sqrt{2} \tilde{x}}^2+(\pp_\a y)^2+ \prt{\sqrt{g_\a''(0)} y}^2~,
\ee
where all the derivatives $\pp_\a$ are with respect to $(\tilde{\t},\tilde{\s})$
\footnote{We use the light-cone space-time coordinates $x^\pm:=x_3\pm x^0,$ and  $x:=x^1+i x^2$.  and without loss of generality we set $x_2=0$ so we do not have to carry the complex norms.}.  We can then identify the bosonic fluctuation spectrum from this Lagrangian where the fields are
\be\la{masses}
\prt{m_{\tilde{x}}^2,m_{\tilde{\phi}}^2,m_y^2}:=\prt{2,4, g_\a''(0)}
\ee
and the bosonic propagator is diagonal. As expected by our construction the square of the mass of the fields $y$, is related to the variable particle mass in the dual particle Hamiltonian description \eq{hamilt}. The factorization of scattering requires all the $2\rightarrow 2$ amplitudes with incoming particles of a certain mass going to different mass vanish. Therefore according to \eq{masses} all amplitudes of an integrable theory must vanish, i.e. $g_\a''(0)=0$, unless $g_\a''(0)=\prt{2,4}$. To constrain the set further it is necessary to compute the amplitudes and the 1-loop corrections to the 2-point functions, and we need to expand the Lagrangian to cubic and the quartic order. The cubic and quartic interaction vertices can be read off the relevant order Lagrangians 
\bea\nn
&&L_3=-4 \tphi\prt{\tx-\tx'}^2+2\tphi\prt{\dot{\tphi}^2-\tphi'^2}+2\tphi\prt{\dot{y}^2-y'^2}\\
&&+ g_a''(0)y^2\prt{ \dot{\tphi}-\tphi'}+\ldots~\la{la3}
\eea
and  
\bea\nn
&&\cL_4= 8\tphi^2\prt{\tx-\tx'}^2+2\tphi^2\prtt{\prt{\pp_\a\tphi}^2+\frac{2}{3}\tphi^2}+2 \tphi^2 \prt{\pp_\a y}^2\\\nn
&&+\frac{g_\a''(0)}{2} y^2\bigg[2\tphi^2+4\tphi \sum_\a \pp_\a \tphi +\prt{\pp_\a\tphi}^2+\prt{\tx+\dot{\tx}}^2\\  \la{la4}
&& +\prt{\tx-\tx'}^2\bigg]+\ldots~.
\eea
All the tree amplitudes turn out to behave qualitatively in the same way, so let us study the tree level contributions of the amplitude $xx\rightarrow yy $ which come from a contact diagram and the s-channel to obtain
\bea\nn
&&\cA_c\prt{xx\rightarrow yy}= g_\a''(0) A_1(p_1,p_2)~,\\ \nn
&&\cA_s\prt{xx\rightarrow yy}=g_\a''(0) A_1(p_1,p_2) A_2(p_1,p_2)~,
\eea
where $A_{1,2}$ are unequal functions of the light-cone momenta of the incoming particles. By considering the total contribution to the amplitude summing both channels and since $A_2(p_1,p_2)$ is a non-trivial function of momenta, the factorization forces $g_\a''(0)$ to be either $0$, so that the amplitude vanishes, or $2$ so that the masses  are equal $m_x=m_y$ and the factorization does not impose any further condition. By summing the contact, the $s$ and $t+u$-channel contributions of the different amplitudes, as in \cite{Wulff:2019tzh},  it turns out that we have already constrained the system enough and the  factorization happens for $g_\a''(0)=0$ where the amplitudes vanish,  or for $g_\a''(0)=2$.  Remarkably, the S-matrix factorization in string picture constrains the spectrum and the theory to the set of values \eq{set}, obtained already by the equivalent particle picture and the Galoisian integrability of the effective particle Hamiltonian \eq{hamilt}.

\noindent \textbf{7. Discussion.}
\lab{sec:s7}
We have formulated an equivalence between the 2-dim $\sigma$-model spectrum expanded on a non-trivial massive vacuum and an effective classical particle Hamiltonian with variable masses and  a non-trivial potential. The equivalence holds for large classes of backgrounds and vacua. We have demonstrated explicitly the formalism in a theory with warp factors of $AdS$ and internal space and a chosen GKP vacuum, where the mass of the excitations of the 2-dim $\s$-model spectrum is dual to the acceleration of the particle's variable mass change. The analytic Galoisian non-integrability on appropriate geodesics of the Hamiltonian system sets an algebraic constraint on the particle masses, such that integrability is allowed. It is remarkable that when this condition is translated to the 2-dim theory through our equivalence, it constrains the spectrum of the 2-dim theory such that the factorization of the $S$-matrix occurs. In particular, the integrability of the classical particle system, implies the factorization of the S-matrix in the dual quantum 2-dim theory! Our proposal initiates the study of analytic non-integrability techniques on the S-matrix vacua in relation to factorization, which goes beyond the classification techniques  used so far.

The particle-string equivalence we have established relies on the fact that the methods of analytic non-integrability and the spectrum of the 2-dim $\s$-model, although independent, based on the same order expansions, that is the reason our formulation applies well to the tree-level factorization. Moreover, our proposal  provides non-trivial evidence beyond any assumptions, on the correlation between integrability and the S-matrix factorization for large class of theories with interactions that may break Lorentz invariance. The (non-)integrability constrains obtained in the particle picture are in agreement in the dual quantum picture with the S-matrix factorization, establishing therefore such connection for a large class of theories. Our methodology provides a ground to develop a formal proof and to clarify explicitly the requirements for the validity of this connection.

\noindent \emph{Acknowledgements.--}
The author acknowledges very useful conversations and correspondence with  I. Antoniadis, J-P. Derendinger, M. Petropoulos, A. Tseytlin, L. Wulff and K. Zoubos. This work is  supported by the Hellenic Foundation for Research and Innovation (HFRI) and the General Secretariat for Research and Technology (GSRT), under
grant agreement No 2344.

\appendix
\textbf{ Appendix A: Differential Galois Group on the Heun Equation.}
The Heun differential equation is brought to its normal form by a transformation $\eta \rightarrow \eta \exp(-\frac{1}{2}\int \k(z) dz )$.
The transformation between the two equations is Liouvillian, so the integrability status between the original and the transformed equations is equivalent. Nevertheless, in general the differential Galois group of the two equations are not the same. In fact the differential Galois group of the transformed normal equation is a subgroup of $SL(2,C)$, while for the initial Heun equation \eq{heunge} this is true if and only if $\k(z)= n f/f'$ with $n\in \mathbb{Z}$ and $f$ is a differentiable function, $f\in K$.  The differential equation in its canonical form has Liouvillian solutions if and only if, its differential Galois group is a proper algebraic subgroup of $SL(2,C)$ \cite{kaplansky1976,moralessl2}. The normal differential equation can be further transformed to the Riccati equation with $g=-\prt{\log \eta}'$.
The criterium for the existence of integrability now takes a more tractable form: The original Heun NVE equation  has Liouvillian solutions if and only if the Riccati NVE equation has an algebraic solution with the degree of the relevant minimal polynomial that belongs to the set $\{1,2,4,6,12\}$. By applying the Kovacic algorithm on the Riccati equation
we find the conditions \eq{criteria}. The Galois differential techniques for such special functions has been also studied in \cite{terab,duval}.

\bibliography{botanyquant}

\end{document}